\pdfoutput=1 

\documentclass[showpacs, pre]{revtex4}   

\usepackage{amsmath}
\usepackage{amssymb}

\usepackage{times}
\usepackage{verbatim}

\usepackage{graphicx}
\usepackage{algorithmic,algorithm}

\DeclareGraphicsExtensions{.pdf}
\begin{document}

\title{Conformity Hinders the Evolution of Cooperation on Scale-Free Networks}

\author{Jorge Pe\~na}
\email{jorge.pena@unil.ch}
\affiliation{Institute of Applied Mathematics, University of Lausanne, Switzerland}

\author{Enea Pestelacci}
\email{enea.pestelacci@unil.ch}
\affiliation{Information Systems Institute, Faculty of Business and Economics, University of Lausanne, Switzerland}

\author{Marco Tomassini}
\email{marco.tomassini@unil.ch}
\affiliation{Information Systems Institute, Faculty of Business and Economics, University of Lausanne, Switzerland}

\author{Henri Volken}
\email{henri.volken@unil.ch}
\affiliation{Institute of Applied Mathematics, University of Lausanne, Switzerland}

\begin{abstract}

\noindent 
We study the effects of conformity, the tendency of humans to imitate locally common behaviors, in the evolution of cooperation when individuals occupy the vertices of a graph and engage in the one-shot Prisoner's Dilemma or the Snowdrift game with their neighbors. Two different graphs are studied: rings (one-dimensional lattices with cyclic boundary conditions) and scale-free networks of the Barab\'asi-Albert type. The proposed evolutionary-graph model is studied both by means of Monte Carlo simulations and an extended pair-approximation technique. We find improved levels of cooperation when evolution is carried on rings and individuals imitate according to both the traditional pay-off bias and a conformist bias. More important, we show that scale-free networks are no longer powerful amplifiers of cooperation when fair amounts of conformity are introduced in the imitation rules of the players. Such weakening of the cooperation-promoting abilities of scale-free networks is the result of a less biased flow of information in scale-free topologies, making hubs more susceptible of being influenced by less-connected neighbors.
\end{abstract}

\pacs{89.75.Hc, 87.23.Ge, 02.50.Le,87.23.Kg}

\maketitle

\section{Introduction}
\label{intro}

Understanding the emergence and stability of cooperation is a central problem in many fields of both natural and social sciences. Researchers have traditionally adopted evolutionary game theory~\cite{Weibull1995} as common formal framework for studying the dynamics of strategy change, and games like the Prisoner's Dilemma (PD) and the Snowdrift Game (SG) as metaphors for the tension between group welfare and individual selfishness. The PD and the SG (also known as Chicken or Hawks-Doves) are two-person, symmetric games in which a given player can be, at each time step, either a Cooperator (C) or a Defector (D). Cs are willing to engage in cooperative tasks, while Ds prefer not to, thus exploiting Cs. 
If two individuals of the same type interact, they both get the reward for mutual cooperation $R$ if they cooperate or the punishment for mutual defection $P$ if they defect. If a D and a C interact, the D receives the temptation to defect $T$ and the C receives the sucker's pay-off $S$. In the PD, the pay-offs are ordered such that $T>R>P>S$ with $2R > T+S$. Since $T>R$ and $P>S$, the only Nash equilibrium of the game is the pure strategy (D,D). 
In this case, the dilemma is caused both by ``greed'' (or the temptation to cheat) and ``fear'' that the other player cheats. In the SG, the order of $P$ and $S$ is reversed, yielding $T>R>S>P$. Thus, when both players defect they get the lowest possible pay-off. The pairs of pure strategies (C,D) and (D,C) are Nash equilibria of the game. There is also a third equilibrium in mixed strategies in which strategy D is played with probability $p$ and strategy C with probability $1-p$, where $p$ depends on the actual
pay-off values. The dilemma in this game is caused only by ``greed'', i.e. players have a strong incentive
to threat their opponent by playing D, which is harmful for both parties if the outcome happens to be (D,D).\\
Conventional evolutionary game theoretical models assume an infinite population in which pairs of randomly drawn individuals interact according to a given game. Selection is strictly pay-off biased, which implies that fitter individuals reproduce more (genetic evolution) or successful individuals tend to be imitated more frequently (cultural evolution). In both genetic and cultural evolution, the evolutionary process can be analytically described by a set of equations called the replicator dynamics~\cite{Weibull1995}. In the SG, the only stable equilibrium of such equations is an internal one, corresponding to the mixed strategy of classical game theory, while the two pure equilibria are unstable.
In the PD, the only stable rest point occurs when the population is entirely composed of Ds: Cs are doomed to extinction in this game.\\
Given these unfavorable predictions for the evolution of cooperation, several mechanisms have been invoked in order to explain why altruism can actually emerge, such as kin selection, group selection, direct reciprocity, indirect reciprocity and network reciprocity~\cite{Nowak2006a}. Network reciprocity~\cite{Nowak1992,Santos2005,Lieberman2005, Szabo2007} arises when individuals occupy the vertices of a graph (modeling spatially subdivided populations or social networks) such that interactions are constrained to direct neighbors. When the population of players possesses such a structure, Cs can survive in clusters of related individuals for certain ranges of the game parameters, as it has been known since the pioneering work by Nowak and May~\cite{Nowak1992}. Among the different conceivable population topologies, scale-free networks have received particular attention since they have been found to promote cooperation to a point that Cs dominate Ds in a significant portion of the parameters' space~\cite{Santos2005}.\\
In addition to positing infinite well-mixed populations, the replicator dynamics relies on the assumption that selection is entirely pay-off biased. Such a premise, although natural to posit in genetic evolution, is less straightforward to postulate in cultural evolution where information is transmitted by means of imitation. Humans not only have a bias for imitating more successful people, but also to conform, or to show a disproportionate tendency to copy the behavior of the majority~\cite{Boyd1985}. Recent empirical research has shown that conformity is an important bias in our social learning psychology~\cite{Coultas2004, Efferson2008}, and that it can partially account for the results obtained in laboratory experiments on social dilemmas~\cite{Carpenter2004, Bardsley2005}. Theoretical research has also shown that conformity can promote cooperation in the PD.  In  the standard case of a large, well-mixed population, the dynamics can lead either to full defection or to bi-stability,
depending on the amount of conformity~\cite{Henrich2001b,Pena2008,Henrich2001a}. In~\cite{Pena2008} the case of square lattices was studied by simulation, with the result that conformity stabilizes cooperation in such population topologies, a result confirmed for rings in~\cite{Mengel2008} and, in a more detailed way, in the work presented here.\\ 
In this paper we investigate the evolution of cooperation when individuals imitate with a given 
amount of conformity and both interaction and imitation are constrained to nearest neighbors in a network. In order to extend previous work~\cite{Nowak1992,Hauert2004,Doebeli2005,Santos2005,tom-luth-giac-06,Pena2008} and to study the influence of the network
topology, we use rings and B\'arabasi-Albert scale-free networks as examples of, respectively, simple degree-homogeneous (i.e. regular) and highly degree-heterogeneous graphs. It will be shown that, while conformity reinforces the cooperation-promoting advantages of network reciprocity in rings, the very same mechanism may strongly hinder the evolution of cooperation when the network topology is scale-free. Indeed, when Cs are not initially in the majority and imitation is partly conformist, scale-free networks are no longer the powerful amplifiers of cooperation expected from the results of previous studies. There is thus an interesting interplay between conformity and network reciprocity so that the cooperation-promoting effects of conformity depend on the particular type of networks on which evolutionary dynamics are played.

\section{Model}
\label{sec:model}

We consider a population of size $N$ where the $i$-th individual is represented by the vertex $v_i$ of an undirected, simple graph $G(V,E)$. The  neighborhood of $i$, $\Gamma(i)$, is the set of all individuals $j$ such that there is an edge $e_{ij} \in E$. The number of neighbors of $i$ is thus the degree $k_i$ of vertex $v_i$.
\begin{figure*}[t]
\centerline{
	\includegraphics[width=5in, bb=0 0 948 687]{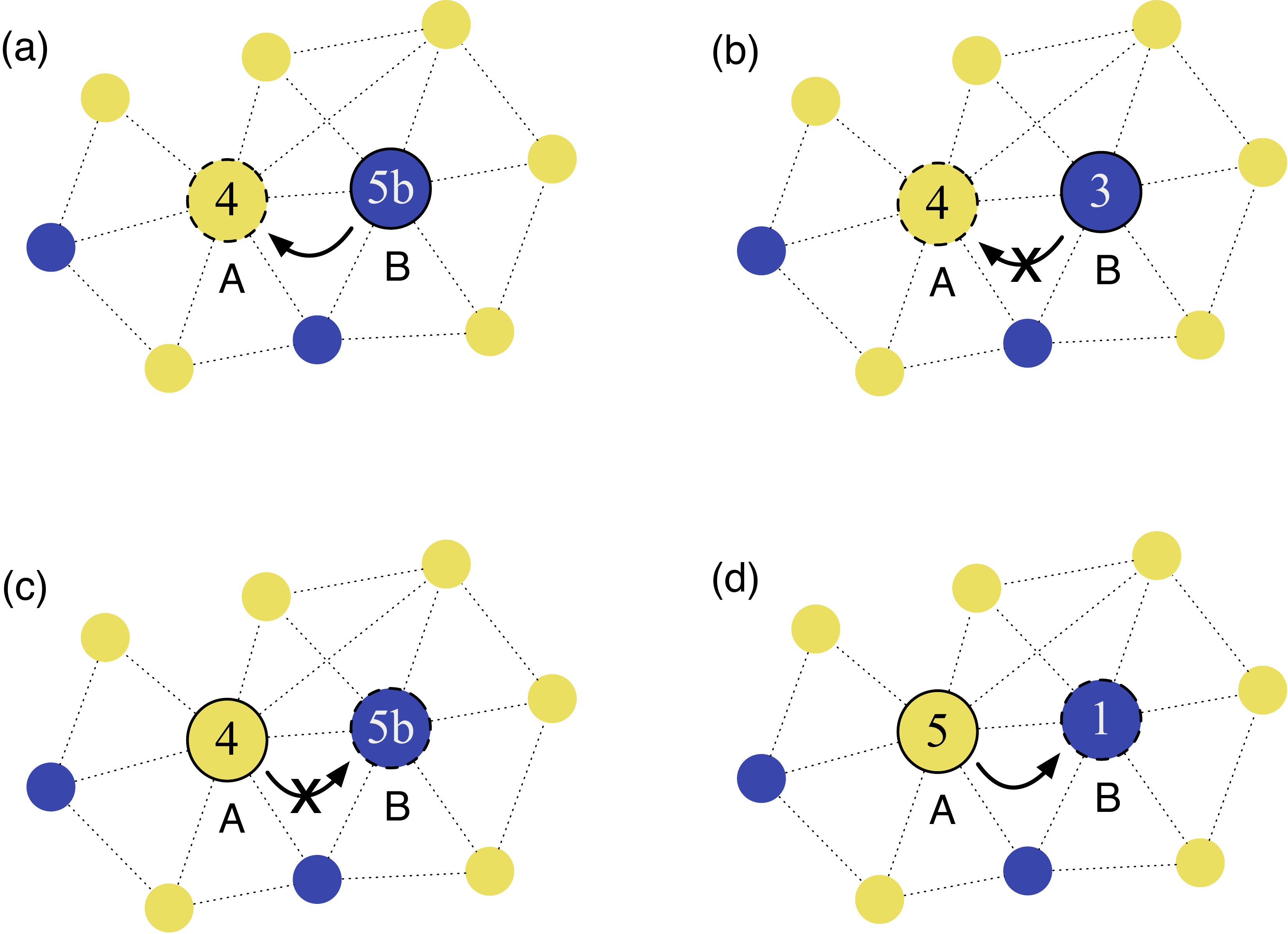}
}

\caption{(Color online) Individuals imitate following two different update rules, each reflecting a different bias of our social learning psychology: pay-off based imitation and conformist imitation. Cooperators are shown in light yellow; defectors in dark blue. Social interaction is modeled by a rescaled Prisoner's Dilemma with $T = b>1$, $R = 1$, $P = S = 0$. (a) Successful pay-off biased transmission. When applying a pay-off biased rule of imitation, $A$ can copy $B$'s strategy and become a defector, since $B$'s pay-off is greater than $A$'s. (b) Unsuccessful conformist transmission. If $A$ were to imitate $B$ according to conformity, no transmission would take place, since defectors are in the minority of $A$'s neighborhood (3 defectors vs. 4 cooperators). (c) Unsuccessful pay-off biased transmission. $B$ will not copy $A$'s strategy under a pay-off biased rule since $A$'s pay-off is smaller than $B$'s. (d) Successful conformist transmission. Conformist transmission from $A$ to $B$ can take place because cooperators constitute the majority in $B$'s neighborhood (5 cooperators vs. 1 defector).}
\label{fig:model}
\end{figure*}
\\
At each time step, each individual is either a C or a D. The system evolves by the successive application of interaction and imitation phases. During the interaction phase, individuals simultaneously engage in a single round of the game with their neighbors. As a result, individual $i$ collects an accumulated payoff $\Pi_i = \sum_{l \in \Gamma(i)} \pi_{il}$, where $\pi_{il}$ is the pay-off player $i$ receives when interacting with player $l$ (e.g. $T$, $R$, $P$ or $S$). During the imitation phase, each individual randomly chooses one of its neighbors as its \textit{cultural model}. Let us denote $i$'s cultural model by $j$. We consider two update rules for the cultural evolutionary dynamics: pay-off biased imitation and conformist imitation. (i) For pay-off biased imitation, $i$ copies $j$'s strategy with a probability given by $f\left((\Pi_j-\Pi_i)/(\theta k_{>})\right)$, where $f(x)$ is equal to $x$ if $x>0$ and $0$ otherwise, $k_{>} = \max \left\{k_i, k_j\right\}$, $\theta=T-S$ in the PD and $\theta=T-P$ in the SG. 
This update rule is a local, finite population analogue of the replicator dynamics, commonly used in the literature~\cite{Hauert2004,Santos2005}. (ii) For conformist imitation the probability that $i$ copies $j$'s strategy is given by $f\left((n_{j|i}-n_{i|i})/k_i\right)$ where $n_{l|i}$ is the number of $i$'s neighbors with strategy $l$. This update rule is related to the majority rule and to the voter model, commonly used in interdisciplinary physics studies~\cite{cox}. In our model individuals imitate according to a pay-off bias with probability $1-\alpha$, and according to a conformist bias with probability $\alpha$. Thus, the parameter $\alpha$ represents the amount of conformity in the individuals' behavior and gives the average proportion of players imitating according to the conformity rule at each time step. When $\alpha=0$ our local dynamics reduce to the strictly pay-off biased imitation rule used in previous studies~\cite{Santos2005,Hauert2004}. Figure~\ref{fig:model} gives some illustrative examples of the imitation dynamics of the proposed model.
\\
In order to allow  comparison with previous studies, we focus on the commonly used rescaled version of 
the PD~\cite{Nowak1992,Santos2005}, for which $T=b$, $1 \le b \le 2$, $R=1$ and $P = S = 0$. The parameter $b$ represents the advantage of defectors over cooperators. For the SG we make, as in~\cite{Santos2005}, $T=\beta>1$, $R = \beta-1/2$, $S=\beta-1$, and $P=0$, such that the cost-to-benefit ratio of mutual cooperation is given by $r=1/(2\beta-1)$. It is worthy of note that, in degree-inhomogeneous networks, the local replicator dynamics using accumulated payoff is not invariant with respect to affine transformations of the payoff matrix~\cite{Tomassini2006a,Luthi-Tom-Pest-09}. Although this fact invalidate generalizations of the obtained results to the extended parameter space, it allows us to compare our results with relevant previous work.\\

Before studying our model with actual network models (rings and scale-free networks) by means of numerical simulation, we briefly present analytical results obtained using the mean-field method and the pair approximation. Such analytical results are important in order to identify the dynamical regions of the system and to serve as starting point for comparisons with the dynamics on actual networks studied in Section~\ref{sec:results}. 

\section{Analytical Results}

\subsection{Mean-Field Approach}

\begin{figure*}[t]
\centerline{
	\includegraphics[width=5.0in, bb=0 0 856 511]{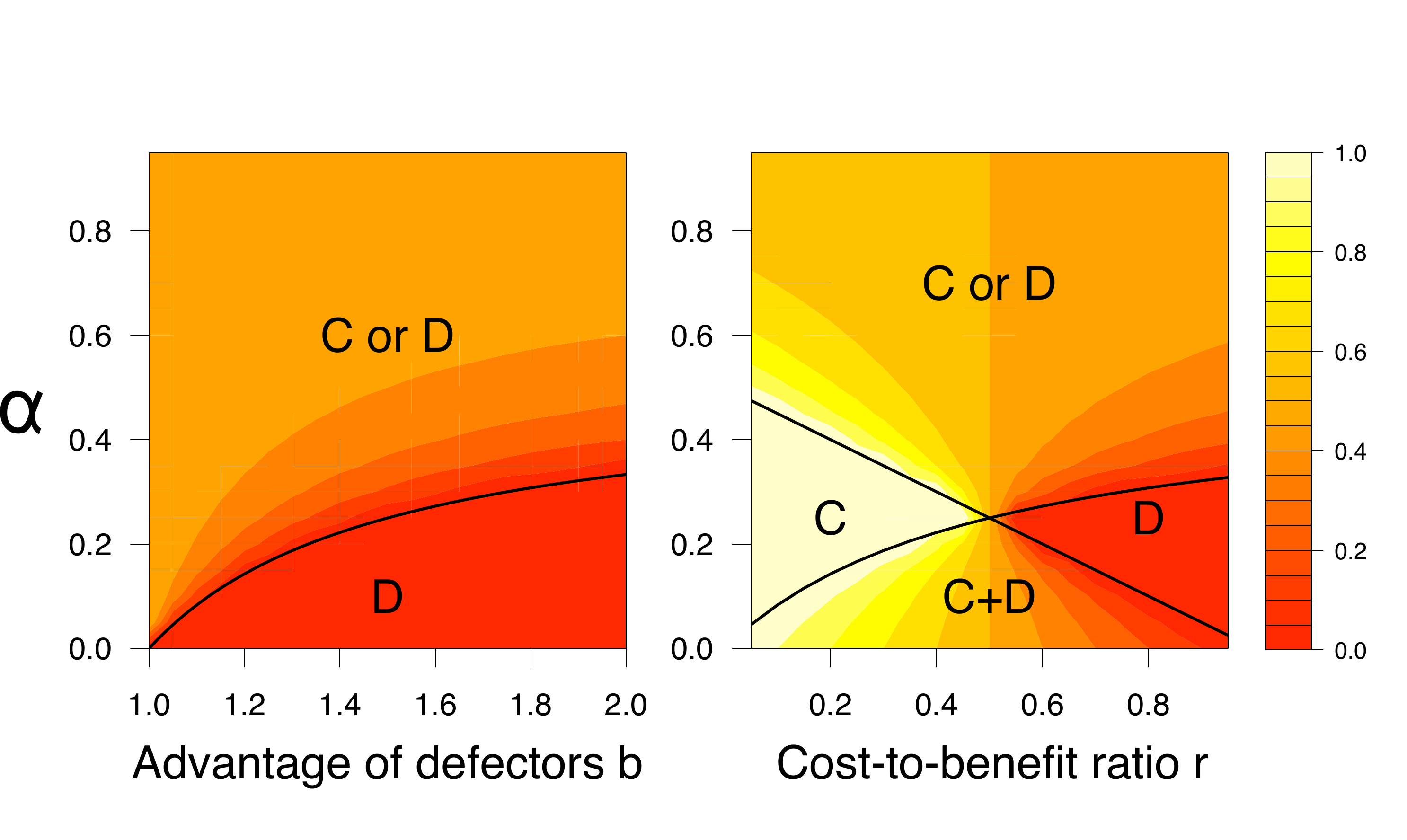}
}
\caption{(Color online) Mean-field solutions of the phase diagrams for the PD with conformity on the $b$-$\alpha$ plane (left) and for the SG with conformity on the $r$-$\alpha$ plane (right). For the PD, the system can be in the dominant defection (\emph{D}) or the bi-stability (\emph{C or D}) regions. For the SG, dominant cooperation (\emph{C}) and co-existence (\emph{C+D}) are also possible outcomes. Darker colors indicate more defection in the average. In the \emph{C or D} region, colors indicate the size of the basin of attraction for the cooperative equilibrium. In the \emph{C+D} region, colors indicate the equilibrium proportion of Cs.}
\label{fig:phasediagramsmf}
\end{figure*}

Within the framework of the traditional mean-field approach~\cite{Szabo2007} network locality is ignored and the system is assumed to have an infinite size, leading to an infinite, well-mixed population. In this case, it is easy to show that the time evolution of the fraction of Cs $\rho$ is ruled by the following equation:
\begin{equation}
\label{eq:modrd}
\dot{\rho} = \rho(1-\rho)\left\{\gamma \left[\pi_C-\pi_D\right] + \alpha(2\rho-1)\right\} ,
\end{equation}
where $\pi_C=\rho R + (1-\rho)S$ and $\pi_D=\rho T + (1-\rho)P$ are the average pay-offs to Cs and Ds, and $\gamma=(1-\alpha)/\theta$. Equation~\ref{eq:modrd} (or a similar formula) has been derived in related work on cultural transmission processes including both pay-off biased imitation and conformist imitation~\cite{Henrich2001a,Henrich2001b,Carpenter2004,Pena2008,Skyrms2005}. The dynamics has the two trivial fixed points $\rho_0^* = 0$ and $\rho_1^* = 1$, as well as (possibly) one internal non-trivial equilibrium given by
\[
\rho^* = \frac{\gamma(P-S)+\alpha}{\gamma\left\{R-T+P-S\right\}+2\alpha} .
\]
For $\alpha = 0$ (pure pay-off biased transmission) Eq.~\ref{eq:modrd} recovers the standard replicator dynamics of the original game, whereas for $\alpha = 1$ (pure conformist transmission), Eq.~\ref{eq:modrd} is equivalent to the replicator dynamics of a pure coordination game with internal (unstable) equilibrium $\rho^*=1/2$. For $0 < \alpha < 1$, variations in the amount of conformity and the entries of the pay-off matrix can change the evolutionary dynamics of the social dilemma. In particular, the global behavior of the system depends on the two critical values $\alpha_D = (S-P)/(\theta+S-P)$ and $\alpha_C = (T-R)/(\theta+T-R)$ so that the system is in one of the following four dynamical regions:
\begin{enumerate}
\item Dominant defection ($\alpha > \alpha_D \wedge \alpha < \alpha_C$): $\rho_0^*=0$ is the only stable equilibrium. In this case, Cs are doomed to extinction regardless of their initial frequency in the population.
\item Co-existence ($\alpha < \alpha_D  \wedge \alpha < \alpha_C$): only the internal equilibrium $\rho^*$ is stable. Cs and Ds coexist in equilibrium at proportions given by $\rho^*$ and $1-\rho^*$, respectively.
\item Bi-stability ($\alpha > \alpha_D \wedge \alpha > \alpha_C$): both $\rho_0^*=0$ and $\rho_1^*=1$ are stable whereas the internal fixed point $\rho^*$ is unstable. In this case, the evolutionary dynamics depends on the initial frequency of Cs, $\rho(0)$. For $\rho(0)>\rho^*$ cooperation prevails, whereas it vanishes for $\rho(0)<\rho^*$.
\item Dominant cooperation ($\alpha < \alpha_D \wedge \alpha > \alpha_C$): $\rho_1^*=1$ is the only stable equilibrium; Cs get fixed regardless of their initial frequency in the population.
\end{enumerate}
These regimes can be seen in Fig.~\ref{fig:phasediagramsmf}, which shows the phase diagrams of the two rescaled games. In the PD with conformity,  $S < P \Rightarrow \alpha_D < 0$, so that only dominant defection and bi-stability are possible. In particular, for the rescaled version of the game, conformity can make the system bi-stable if $\alpha > (b-1)/(2b-1)$. However, for all values of $b$ in the bi-stability region, the basin of attraction of $\rho_0^*$ is greater than the basin of attraction of $\rho_1^*$, i.e. Cs initially in the minority are doomed to extinction regardless of their initial proportion and the values of $b$ and $\alpha$. In the SG with conformity, the four dynamical regions above described are possible, with $\alpha_D = (1-r)/2$ and $\alpha_C = r/(1+2r)$. In the co-existence region, the equilibrium proportion of Cs is larger than what is expected in the $\alpha=0$ case when $r<1/2$ and smaller when $r>1/2$. In the bi-stability region, the basin of attraction of $\rho_1^*$ is greater than the basin of attraction of $\rho_0^*$ for $r<1/2$. 

In sum, conformity can promote cooperation in the PD to a certain degree in the mean-field limit. If in the majority (and if conformity is strong enough) Cs now have a chance of surviving invasion from Ds, and eventually take over the whole population~\cite{Pena2008}. In the SG, whether conformity helps or hinders the evolution of cooperation actually depends on the cost-to-benefit ratio $r$. Cs are favored for $r<1/2$ and disfavored for $r>1/2$.

\subsection{Pair Approximation}
\label{pa}

Pair approximation~\cite{Matsuda1992,Baalen1998} improves over traditional mean-field approach for structured populations by considering the frequency of strategy pairs (i.e. C-C, C-D and D-D). Since the technique assumes regular graphs without loops, it only applies to Bethe lattices in a strict sense~\cite{Hauert2005}.  However, pair approximation has been used to predict evolutionary dynamics on more general regular graphs with considerable success~\cite{Hauert2004,Hauert2005}. We extended the pair-dynamics model presented in the Supplementary Information of Ref.~\cite{Hauert2004} to investigate the cultural evolutionary dynamics of social dilemmas on graphs.
The pair approximation of our model leads to a system of ordinary differential equations tracking changes in the proportions $p_{c,c}$, $p_{c,d}$ and $p_{d,d}$ of, respectively, the C-C, C-D and D-D links in the population graph. The resulting system, although impossible to solve analytically due to the nonlinearity of the equations, can be solved numerically after specifying suitable initial conditions.
\begin{figure*}[t]
\centerline{
	\includegraphics[width=5.0in,bb=0 0 1039 1011]{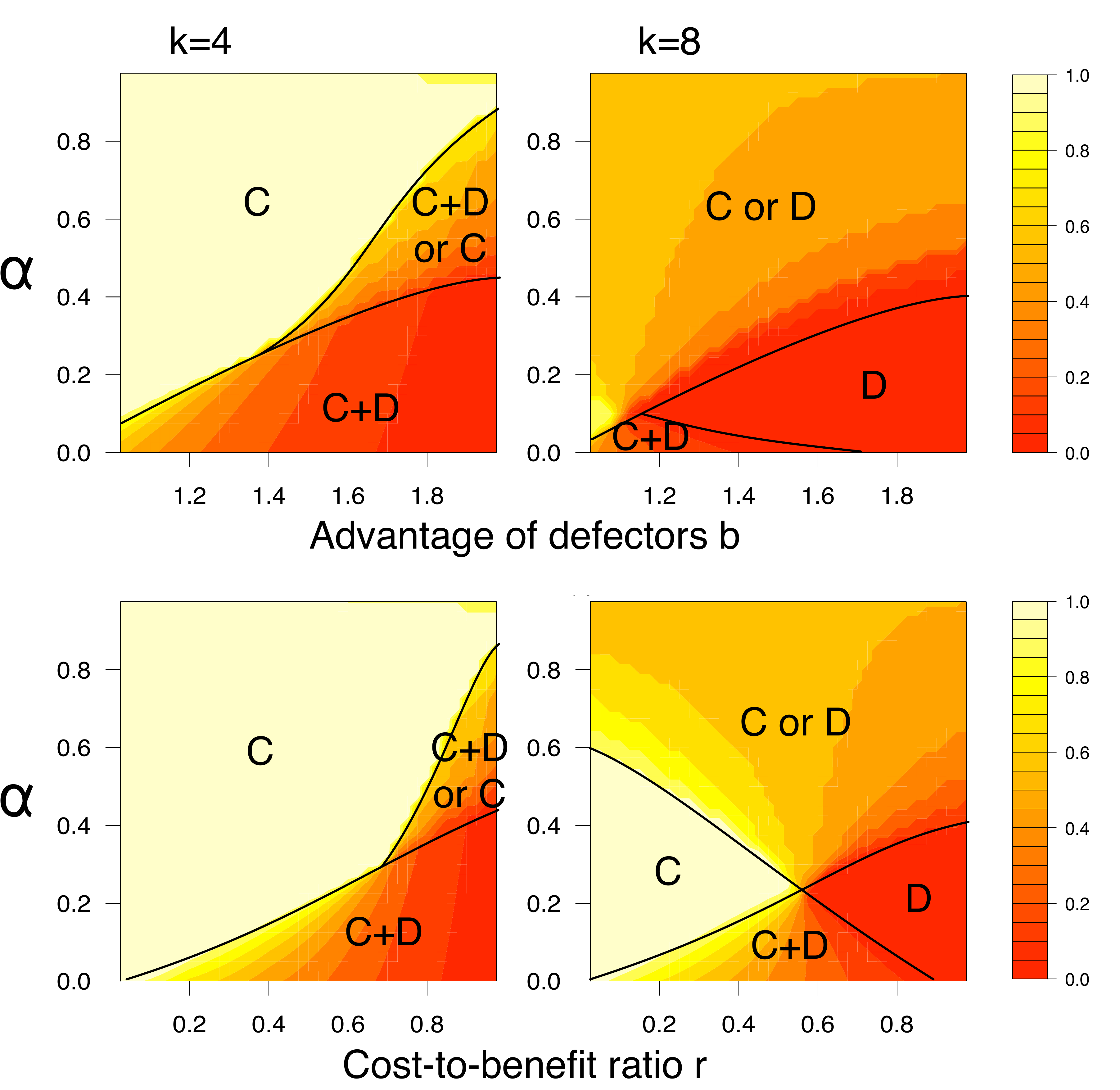}
}
\caption{(Color online) Pair approximations of the phase diagrams for the PD with conformity on the $b$-$\alpha$ plane (top row) and the SG with conformity on the $r$-$\alpha$ plane (bottom row). The first column shows the results for $k=4$, the second column for $k=8$. The system exhibits different dynamical regimes depending on the game: dominant cooperation (\emph{C}), dominant defection (\emph{D}), coexistence (\emph{C+D}), and bi-stability (\emph{C or D} and \emph{C+D or C}). In the \emph{C+D or C} region, the system can stabilize in a mixed state or in pure cooperation. Darker colors indicate more defection in the average.}
\label{fig:phasediagramspa}
\end{figure*}
\\
Figure~\ref{fig:phasediagramspa} shows the phase diagrams for the pair approximation of our model, for regular graphs with degree $k=4$ and $k=8$. The figures were constructed by numerically integrating the equations under different initial proportions of Cs ($\rho(0)=\{0.1,0.2, \ldots,0.9\}$) and averaging over all initial conditions. Pure spatial effects can be seen when $\alpha=0$. For the PD, the dynamical regime of the game is no longer of dominant defection, but of co-existence. Locality of interactions thus favors Cs by allowing them to survive extinction. In addition to this classical result, for $k=4$ conformity is largely favorable to Cs. Indeed, augmenting $\alpha$ increases the proportion of Cs in the co-existence region and, depending on the value of $b$, can shift the system to the region of dominant cooperation. In the SG with $k=4$ conformity has similar effects, resulting in an analogous dynamic picture. The fact that the SG represents a less stringent dilemma makes larger the area of dominant cooperation. For $k=8$, phase diagrams get closer to those predicted by the mean-field method (see Fig.~\ref{fig:phasediagramsmf}) but important levels of cooperation are still sustained. In the PD, for instance, the basins of attraction of the cooperative equilibrium in the bi-stability region are larger than those expected in a well-mixed population (compare the top right panel of Fig.~\ref{fig:phasediagramspa} with the left panel of Fig.~\ref{fig:phasediagramsmf}).
\\
In a nutshell, when the population of players possesses local structure, a given amount of conformity in the imitation rules of the players is able to foster cooperation, at least for low values of the mean degree $k$. The reason for this is the easier formation of clusters of individuals playing the same strategy induced by conformist imitation.

\section{Simulation Results}
\label{sec:results}

We now turn our attention to actual networks as population topologies, in particular (i)  
rings (regular 1D-lattices with cyclic boundary conditions) with degrees $k=4$, $k=8$ and $k=16$, and (ii) Barab\'asi--Albert scale-free networks~\cite{alb-baraba-02} with average degrees $\bar k=4$, $\bar k=8$, and $\bar k=16$. For both types of networks we generated graphs of size $N=10^4$. In the case of rings, graphs are constructed by arranging the nodes on a circle and connecting each node to the $k$ most-neighboring nodes.
\\
We study the model by Monte Carlo simulations in populations randomly initialized with $50\%$ Cs and $50\%$ Ds (but see Section~\ref{sub:initial} for other initial conditions). The probability $\alpha$ of conformist transmission was set to $\alpha \in [0,0.5]$ in steps of $0.1$. We privilege values of $\alpha \leq 0.5$ so that dynamics are primarily driven by pay-off differences in the competing strategies. However, we also study the limiting case $\alpha=1$ in Section~\ref{sub:conformist} and the case $0 \leq \alpha \leq 1$ in Section~\ref{sub:initial}. The advantage of defectors $b$ (PD) and the cost-to-benefit ratio $r$ (SG) were varied in steps of $0.05$. We carried out 50 runs for each couple of values of $\alpha$ and the game parameter. For the scale-free networks, we used a fresh graph realization in each run. The average final frequency of Cs $\hat{\rho}$ was obtained by averaging over $10^3$ time steps after a relaxation time of $10^4$ time steps.

\begin{figure*}[t]
\centerline{
	\includegraphics[width=4.5in,bb=0 0 1065 930]{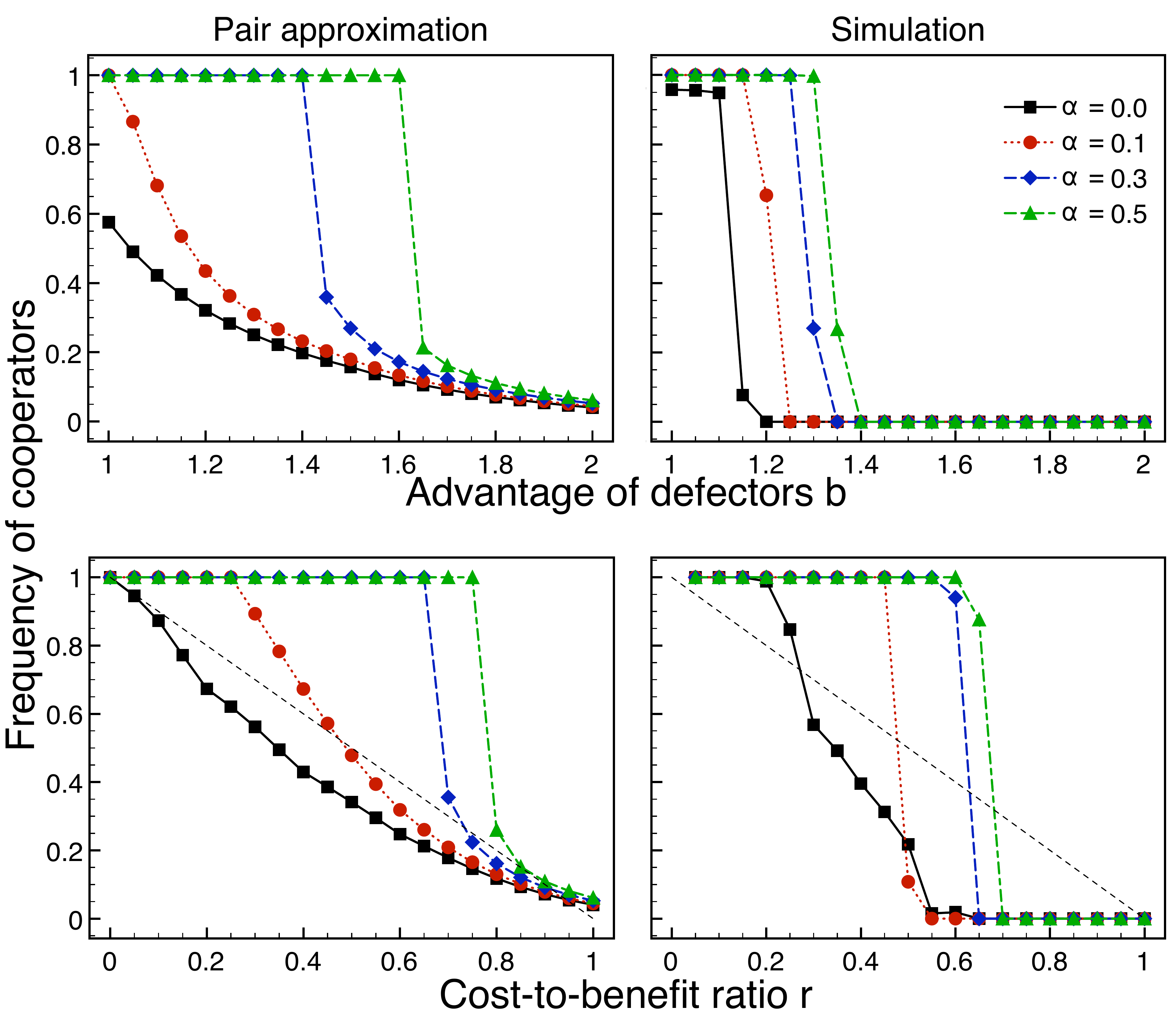}
}
\caption{(Color online) Final average frequency of Cs on rings with $k=4$ for the PD (upper panels) and the SG (lower panels) as a function of $b$ or $r$ for different values of the propensity to conform $\alpha$. Results by Monte Carlo simulations are shown in the right panels while predictions by pair approximation are shown in the left panels. Mean-field approximations for the SG and $\alpha = 0$ are shown with dotted lines.}
\label{ring_k4}
\end{figure*}
\begin{figure*}[t]
\centerline{
	\includegraphics[width=4.5in,bb=0 0 1065 930]{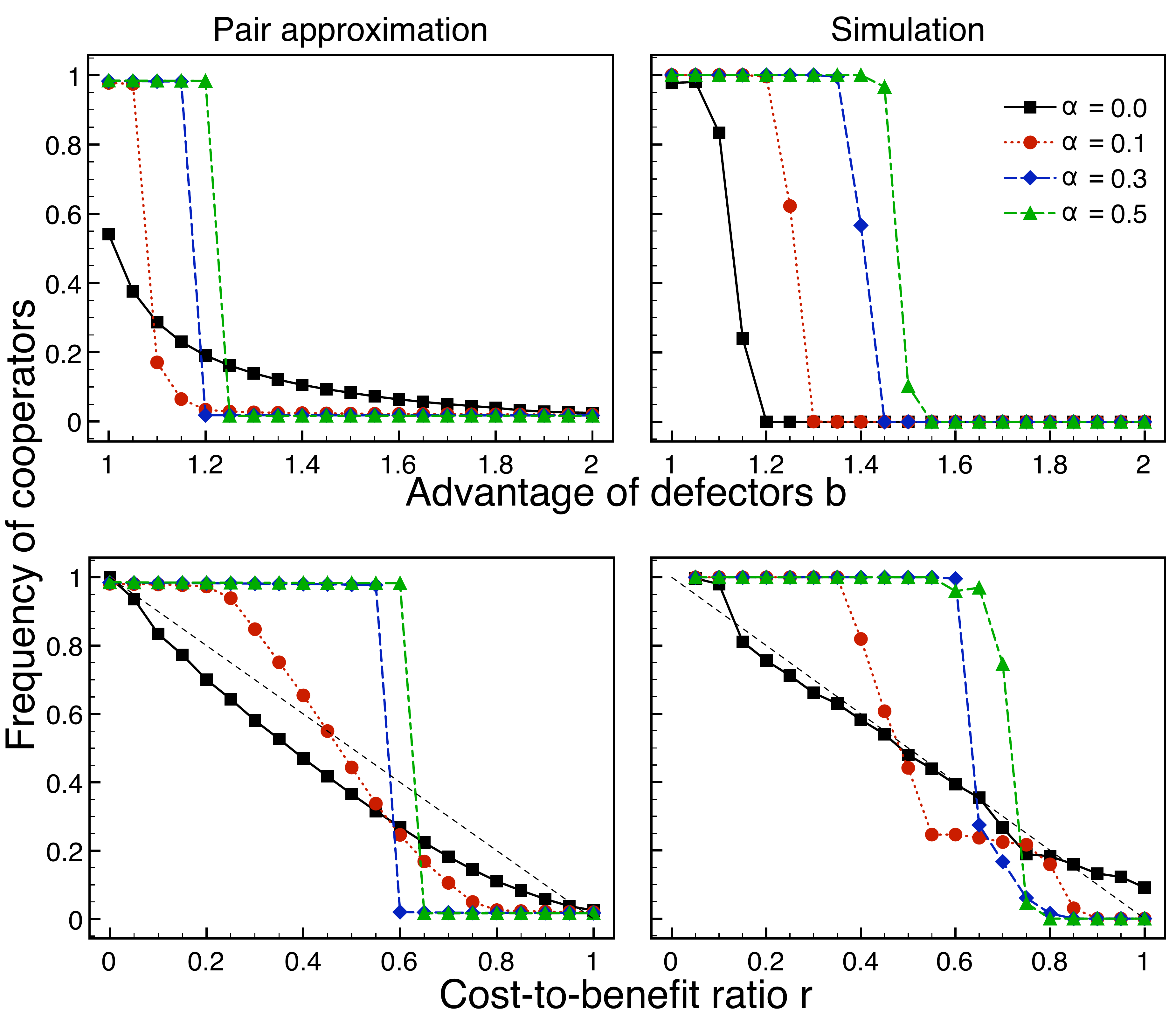}
}

\caption{(Color online) Final average frequency of Cs on rings with $k=8$ for the PD (upper panels) and the SG (lower panels) as a function of $b$ or $r$ for different values of the propensity to conform $\alpha$. Results by Monte Carlo simulations are shown in the right panels while predictions by pair approximation are shown in the left panels. Mean-field approximations for the SG and $\alpha = 0$ are shown with dotted lines.}
\label{ring_k8}
\end{figure*}

\subsection{Results for Rings}

Figure~\ref{ring_k4} summarizes the results obtained for the PD and the SG on rings with $k=4$. These plots confirm the results previously obtained for the standard $\alpha=0$ case on these population topologies~\cite{Santos2005,tom-luth-giac-06}, which in turn are qualitatively similar to those obtained for square lattices~\cite{Nowak1992,Hauert2004}.  In the PD, Cs are able to survive for low values of $b$ by forming clusters wherein they interact more often with their own strategy than what is expected in well-mixed populations. Cs can thus benefit from mutual cooperation and counterbalance the exploitation of Ds at the borders of the clusters~\cite{Doebeli2005}. In the SG, spatial structure hinders the evolution of cooperation~\cite{Hauert2004}, such that only for small values of $r$ (i.e. $r<0.3$) the final fraction of Cs is higher than what is expected in a well-mixed population. As it is evident from our results, conformity enhances cooperation in rings, moving rightward the critical value $b^*$ for which $\hat{\rho}=0$ in the PD, and the value $r^*$ for which the $\hat{\rho}$ becomes smaller than the corresponding proportion in a well-mixed population in the SG. Furthermore, the different curves are ordered in a way that the higher $\alpha$, the higher $\hat{\rho}$ for all values of $b$ and $r$ (except for the SG, $r=0.5$, $\alpha=0.1$) and the larger the critical values $b^*$ and $r^*$.\\
Figure~\ref{ring_k8} plots the results for rings with $k=8$. In the PD, conformity enhances cooperation even more pronouncedly than in the $k=4$ case. Indeed, the threshold $b^*$ has moved rightward for every value of $\alpha$. Such trend is still present in the results obtained for rings with $k=16$ (not shown here to avoid cluttering the figures). In the SG, the increase in the degree of the graph makes conformity cooperation-enhancing up to a threshold value $\hat r$ (where a curve with $\alpha > 0$ crosses the curve with $\alpha=0$) but detrimental afterwards. As $b^*$ in the PD, also $\hat r$ moves rightward as $\alpha$ increases.\\
With respect to simulation results, pair approximation tends to underestimate cooperation for low values of $\alpha$ and $b$ or $r$ and to overestimate it for medium to large values of these parameters. For the PD with conformity, results for $k=8$ are rather pessimistic and are much closer to what we have obtained for random graphs (data not shown here). This is not surprising since random graphs are locally similar to Bethe lattices~\cite{Bollobas1995}. Notice, however, that pair approximation predicts reasonably well the cooperation-enhancing effects of conformity in the PD and the ordering of the curves for different values of $\alpha$. Also, for the SG, pair approximation accurately predicts the fact that the curves with conformity ($\alpha > 0$) are above the curve without conformity ($\alpha = 0$) when $k=4$ (Fig.~\ref{ring_k4}, lower panels), but that they cross it when $k=8$ (Fig.~\ref{ring_k8}, lower panels). This means that pair approximation correctly predicts the fact that, for $k=8$, there is a point up to which conformity helps Cs but beyond which Ds are favored with respect to the standard case without conformity.

\subsection{Results for Scale-Free Graphs}
\label{rsf}
\begin{figure*}[t]
\centerline{
	\includegraphics[width=4.5in,bb=0 0 1065 930]{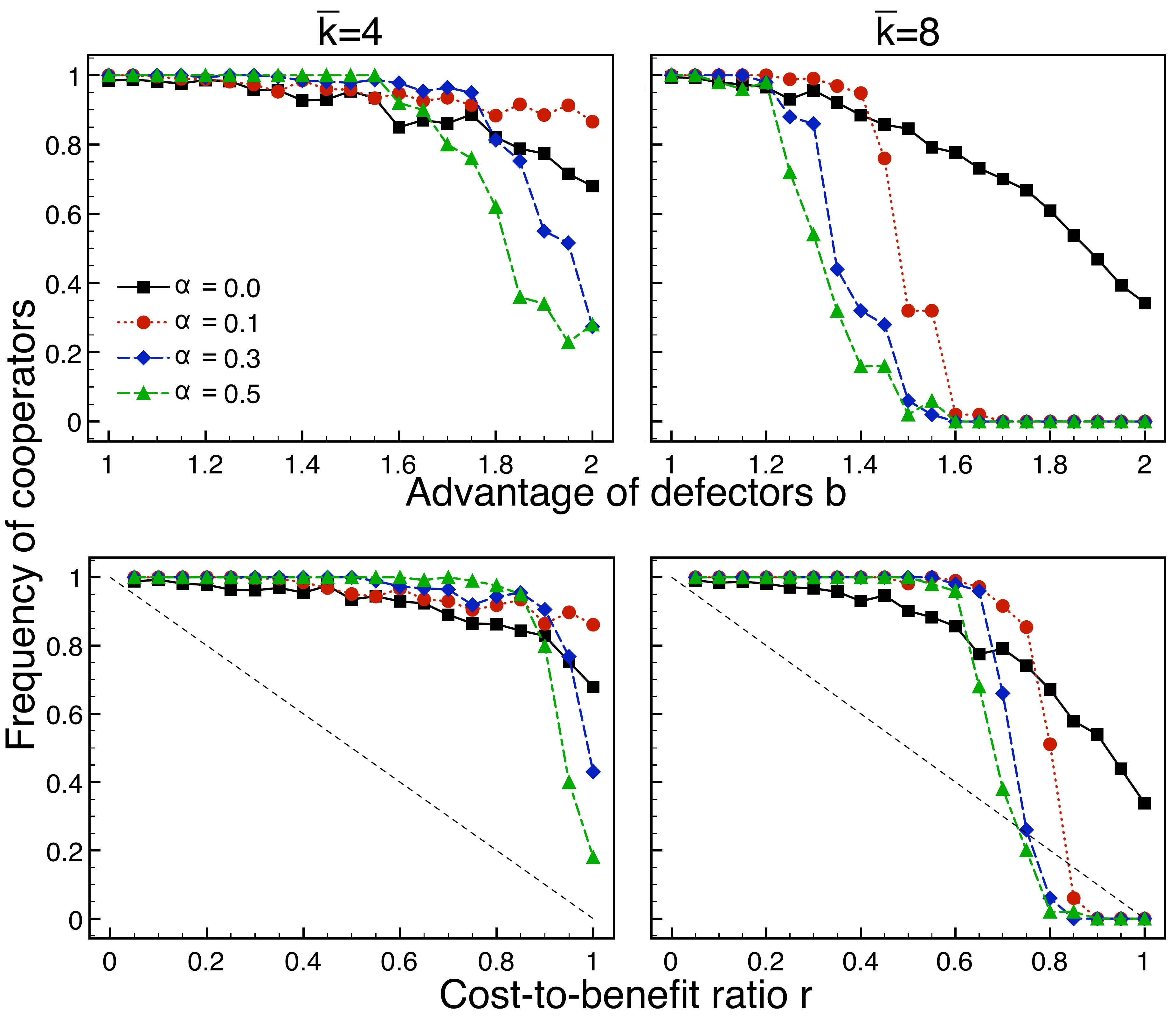}
}
\caption{(Color online) Final average frequency of Cs on scale-free networks for the PD (upper panels) and the SG (lower panels) as a function of $b$ (PD) or $r$ (SG) for different values of the propensity to conform $\alpha$. Results are shown for $\bar k=4$ (left panels) and $\bar k = 8$ (right panels). Mean-field approximations for the SG and $\alpha = 0$ are shown with dotted lines.}
\label{fig:sf}
\end{figure*}
Let us now turn our attention to the results obtained for scale-free networks (Fig.~\ref{fig:sf}). When imitation is strictly pay-off biased ($\alpha=0$) these degree-heterogeneous graphs importantly foster cooperation in both the PD and the SG with respect to what is obtained in rings and other degree-homogeneous graphs~\cite{Santos2005}. As an aside, we note that the higher the average degree $\bar k$, the lower the gains in cooperation~\footnote{When comparing our results with those of~\cite{Santos2005}, note that the curves are in the wrong order in~\cite{Santos2005} as cooperation should decrease with increasing mean degree for scale-free networks.}.
The addition of conformity has important consequences in the evolution of cooperation on scale-free graphs. In the PD, conformity improves $\hat{\rho}$ for all values of $b$ only for a scale-free topology with $\bar k = 4$ and $\alpha < 0.3$. For the other cases, conformity does not hamper cooperation for small values of $b$ but is detrimental for medium to large values of the game parameter. Furthermore, the threshold value $\hat b$ above which $\hat{\rho}$ is higher than in the case without conformity is a monotonically decreasing function of both $\alpha$ and $\bar k$, such that the higher the amount of conformity and the average connectivity of the graph, the smaller the value of $\hat b$.
Particularly, for scale-free networks with $\bar k = 8$ and $\alpha \geq 0.2$, conformity weakens the advantage of these graphs in promoting cooperation to a point that $\hat{\rho}$ becomes comparable to the corresponding fraction obtained in rings (compare the right upper panels of Fig.~\ref{fig:sf} and Fig.~\ref{ring_k8}). \\
Results for the SG on scale-free networks (lower panels of Fig.~\ref{fig:sf}) are qualitatively similar to those obtained for the PD. Again, conformity is beneficial for cooperation for all values of the game parameter $r$ only for $\bar k = 4$ and $\alpha < 0.3$. For the remaining cases, there is a threshold value $\hat r$ of the cost-to-benefit ratio above which $\hat{\rho}$ is smaller than the corresponding frequency of Cs in the $\alpha=0$ case. We note again the fact that the higher the value of $\alpha$, the lower the value of $\hat r$. Finally, and as in the PD, for $\bar{k}=8$ and $\alpha \geq 0.2$ there are no important quantitative differences in $\hat{\rho}$ between rings and scale-free networks: scale-free networks have again lost the cooperation-enhancing capabilities they feature when imitation is strictly pay-off biased. For $\bar k = 8$ and high values of $r$, the addition of conformity can even make Cs go extinct, which would not happen in the non-conformist case.

\begin{figure*}[t]
\centerline{
	\includegraphics[width=7.2in,bb=0 0 2170 960]{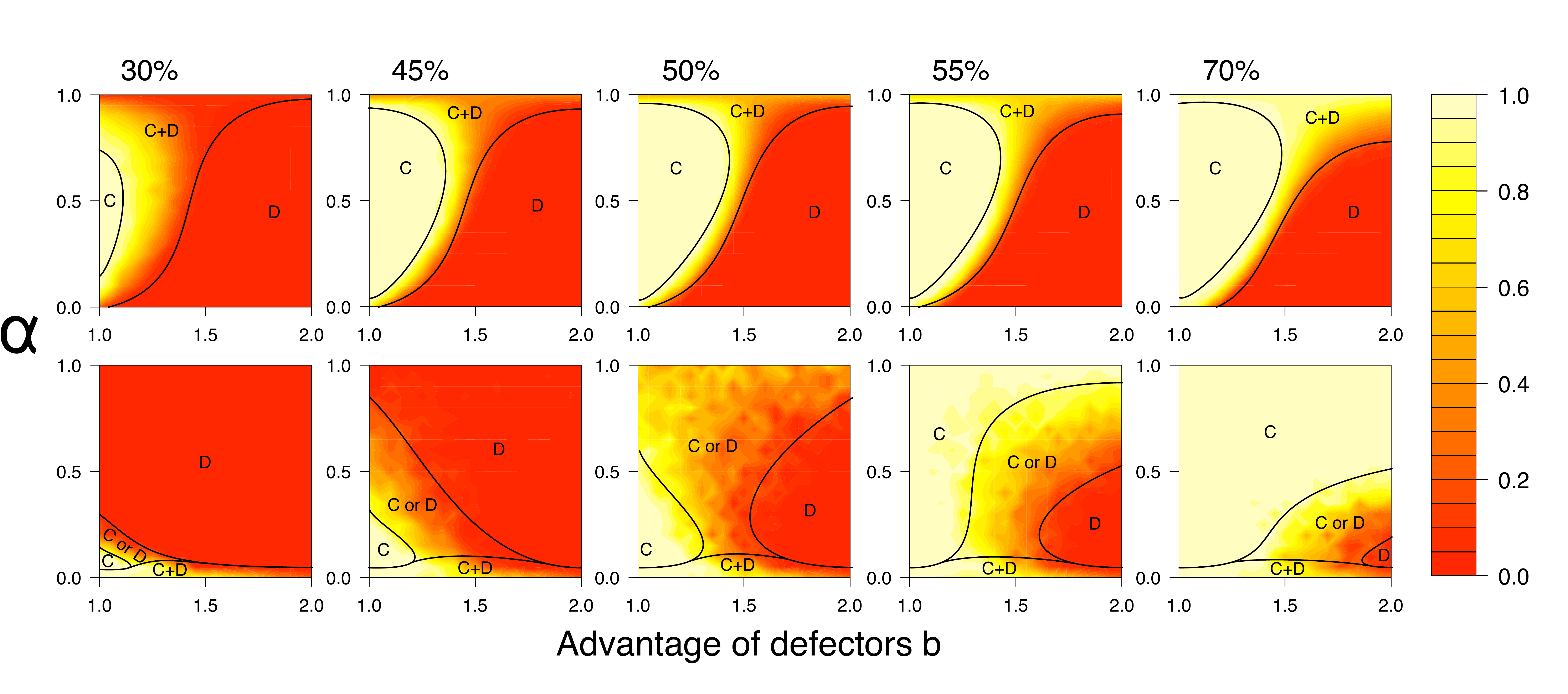}
}
\caption{(Color online) Phase diagrams for the PD game on rings with $k=8$ (top row)
and for scale-free graphs with $\bar k=8$ (bottom row) as a function of  $b$ and $\alpha$.
The images are for increasing initial fractions of cooperation $\rho(0)$ from left to right.}
\label{ratio}
\end{figure*}
\subsection{Dependence on the initial conditions}
\label{sub:initial}

In order to investigate the robustness of cooperation and to study the influence of the initial fraction of Cs $\rho(0)$ we have also run simulations for the PD on rings and scale-free graphs for $\bar k=8$ starting from values of $\rho(0)$ other than $0.5$, and on an extended range of values of $\alpha$ going from $0$ to $1$. Results are shown in Fig.~\ref{ratio} in the form of phase diagrams for each initial condition. In contrast to the notion of bi-stability in a system of ordinary differential equations (such as those resulting from the mean-field approach and the pair approximation), here we define bi-stability as the ability of the system to reach either full cooperation or full defection starting from the same \emph{global} initial conditions, due to its stochastic dynamics and finite size.
\\
Not unexpectedly, initial conditions influence the final outcomes of the simulations, so that the strategy initially in the majority is always favored with respect to the case when $\rho(0)=0.5$. Notice, however, that the effects of conformity are still qualitatively different for each of the two types of networks considered in this study. On these phase diagrams the transition from the region of dominant cooperation (\emph{C}) to dominant defection (\emph{D}) is steeper on rings,  where the two zones with monomorphic populations are divided by a narrow region of co-existence (\emph{C+D}). On scale-free networks a large region of bi-stability (\emph{C or D}) tends to be formed in the middle of the parameter's space, being the largest for $\rho(0)$ close to $50\%$. Indeed, the cultural evolutionary dynamics are much more sensitive to the initial conditions when applied on top of scale-free networks than when they are played on top of rings. For rings, conformity favor Cs even if they are initially in the minority, such that, in general, the higher the value of $\alpha$ the higher the final fraction of Cs in the population. For scale-free networks, conformity can be favorable to cooperation when Cs are initially in the majority, but decidedly detrimental if they are in the minority. The remarkable observation is that in scale-free networks even a small change in the initial fraction of Cs can drastically change the final outcome (see the second and fourth images in the lower row of Fig.~\ref{ratio} for $\rho(0) = 0.45$ and $\rho(0)=0.55$). It would be tempting to compare the numerical results for scale-free graphs with those obtained analytically in the mean-field case and with the pair approximation (Figs.~\ref{fig:phasediagramsmf} and~\ref{fig:phasediagramspa}). However, this cannot be done as both the mean-field and pair approximation approaches give poor results in highly degree-inhomogeneous networks.
\begin{figure*}[t]
\centerline{
	\includegraphics[width=4.5in,bb=0 0 1680 832]{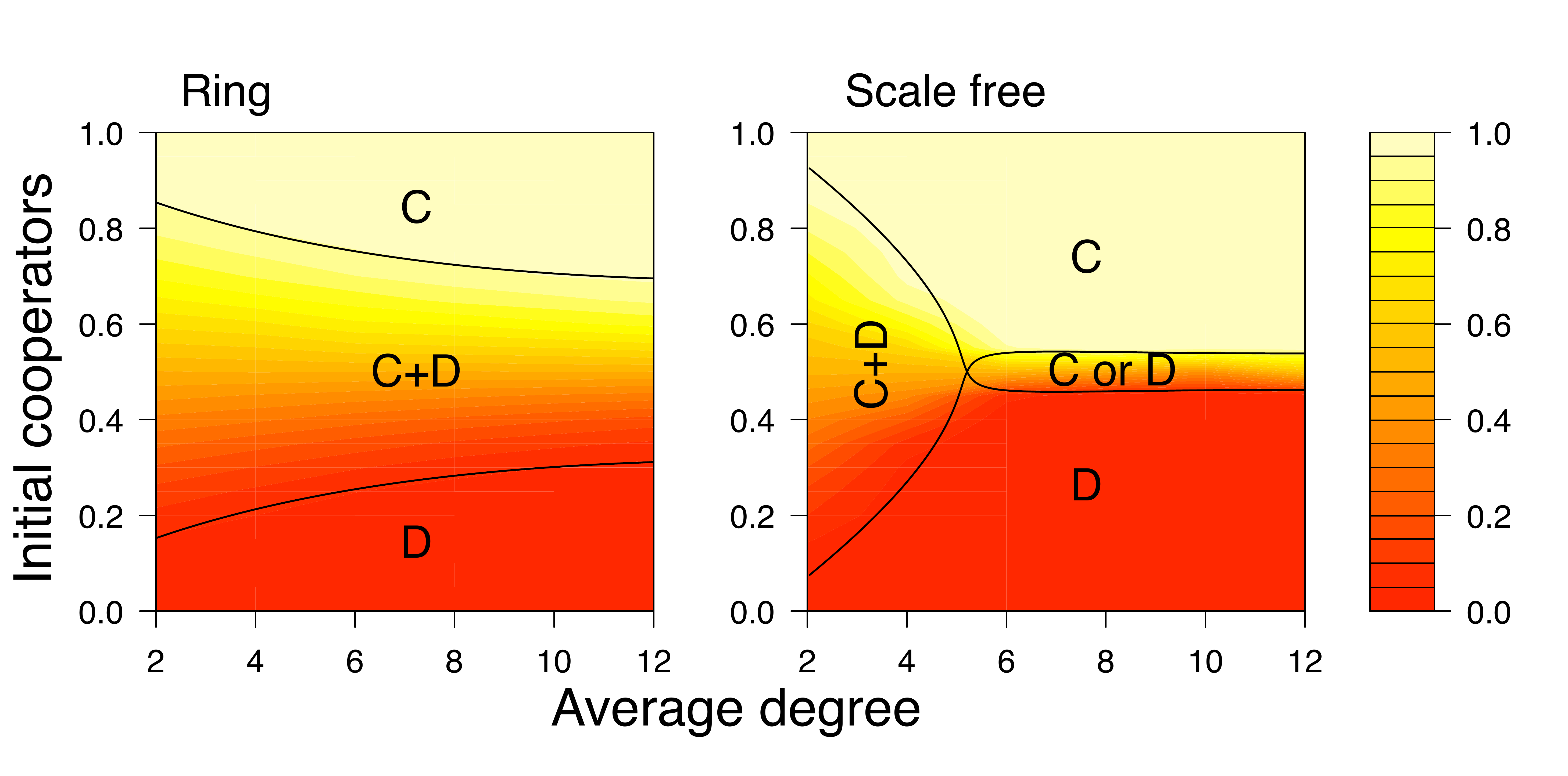}
}
\caption{(Color online) Final population composition as a function of the average
degree $\bar k$ and the initial proportion of cooperators when imitation is purely conformist  ($\alpha=1$).}
\label{alpha_1}
\end{figure*}

\subsection{Pure conformist dynamics}
\label{sub:conformist}

We briefly comment on the case with $\alpha=1$ which is special as the dynamics is completely driven by the majority rule and games' payoffs play no role. Figure~\ref{alpha_1} shows what happens in this case as a function of the network's average degree $\bar k$ and the initial proportion of Cs. 
For $\bar k = 2$ there is a large co-existence region for both graphs, and the pure equilibria have relatively small basins of attraction. With increasing $\bar k$, the co-existence region decreases so that a greater connectivity favors fixation in a monomorphic population. Whereas in rings co-existence is still reached for $\bar k$ as large as 12, for scale-free networks such regime disappears for $\bar k > 5$. For these networks, only in the narrow central strip around $\rho(0)=0.5$ may bi-stability arise. Note that in this case the C and D labels indicating cooperators or defectors are purely conventional as payoffs (and so, the behavioral strategies of the individuals) are completely ignored.

\section{Discussion}
\label{sec:disc}

\begin{figure}[t]
\centerline{
	\includegraphics[width=3.4in,bb=0 0 850 800]{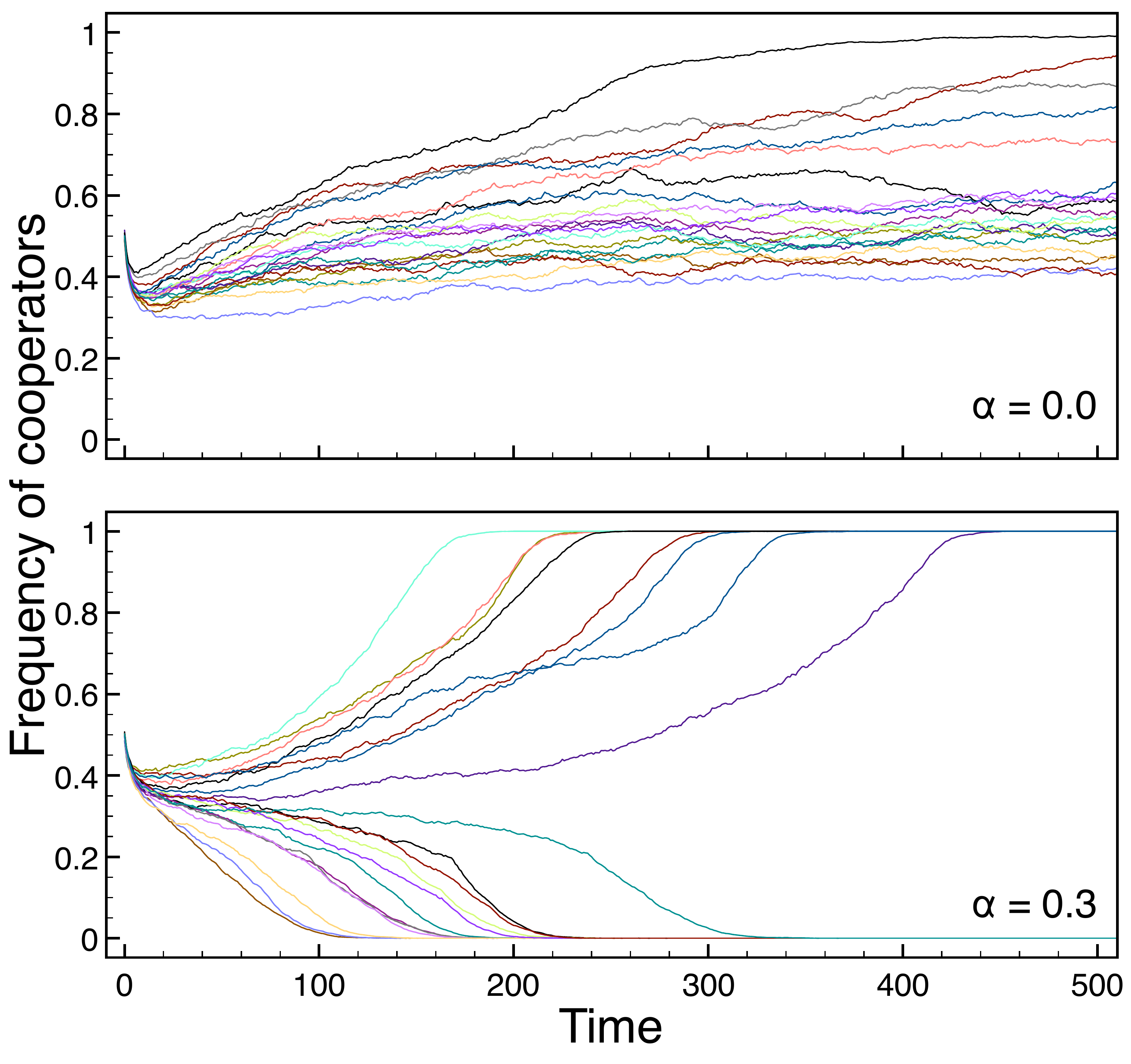}
}
\caption{(Color online) Evolution of the frequency of Cs on scale-free networks ($\bar k = 8$) during the first 500 time steps for the PD, without conformity (upper panel) and with conformity (lower panel). In both figures, $b=1.35$. 20 distinct curves are shown.}
\label{dyn_sf}
\end{figure}

Conformity and network reciprocity are able to act together and foster cooperation in degree-homogeneous graphs for social dilemmas such as the PD and the SG. The basic principle behind network reciprocity is the formation of clusters of related individuals leading to assortative interactions that favor Cs. Conformity further helps such cluster formation thus improving the efficiency of cooperative behavior in a network of interacting individuals.\\
More interestingly, conformity may hinder the evolution of cooperation on the otherwise cooperation-promoting scale-free networks. The different dynamical organization of cooperation in degree-heterogeneous graphs with conformity can explain the reason of such phenomenon. When individuals imitate exclusively according to a pay-off bias, Cs and Ds coexist in quasi-equilibrium, with some nodes fixed in cooperative or defective behavior and others where there is no fixation and cycles of invasion follow indefinitely~\cite{Gomez-Gardenes2007a}. Thus, the gradual drop in cooperation seen in Fig.~\ref{fig:sf} for the case $\alpha=0$ is mostly due to fluctuating individuals spending less and less time engaging in cooperative behavior. This dynamical picture changes when individuals imitate not only according to a pay-off bias, but also to conformity. In this case, for $\bar k = 8$, the population always reaches one of the two absorbing states, so that in the limit only one strategy gets fixed: Cs for low values of $b$, Cs or Ds (with a certain probability) for intermediate values of $b$, and Ds for large values of $b$ (see also the bottom row of Fig.~\ref{ratio}). In general, and contrary to what happens without conformity, intermediate levels of cooperation for $\alpha>0$ (when averaging over several runs) are not the result of the co-existence or fluctuation of different strategies but of the fact that, for an interval of values of $b$, whose length increases with $\alpha$, the system sometimes converges to the cooperative equilibrium and some others to the defective equilibrium (see Fig.~\ref{ratio} bottom row,
central image). Additionally, evolutionary dynamics develop much faster in the presence of conformity. Figure~\ref{dyn_sf} illustrates these observations for the case of scale-free networks with $\bar k = 8$ and $b=1.35$. Without conformity (upper panel of Fig.~\ref{dyn_sf}) the fraction of Cs for each run slowly increases during the initial part of the simulation until, eventually, it stabilizes around 0.9. Conversely, with conformity (lower panel of Fig.~\ref{dyn_sf}), very early in the evolutionary process the population goes either to full cooperation or to full defection.\\
We can gain an insight into the interplay between network reciprocity and conformity by making use of the notion of the temperature of players~\cite{Lieberman2005, Masuda2007}. Hot players are those who play more since they have a large number of neighbors, whereas cold players are those who have few neighbors and, consequently, play less games. By playing more often, and provided that pay-offs are positively biased (i.e. $S\geq0$ in the PD), hot players get higher accumulated payoffs than cold players. Under pure pay-off biased imitation ($\alpha=0$) this implies that hot players are also more successful in being imitated and in disseminating their strategies~\cite{Masuda2007}.
\begin{figure}[t]
\centerline{
	\includegraphics[width=3.4in,bb=0 0 850 800]{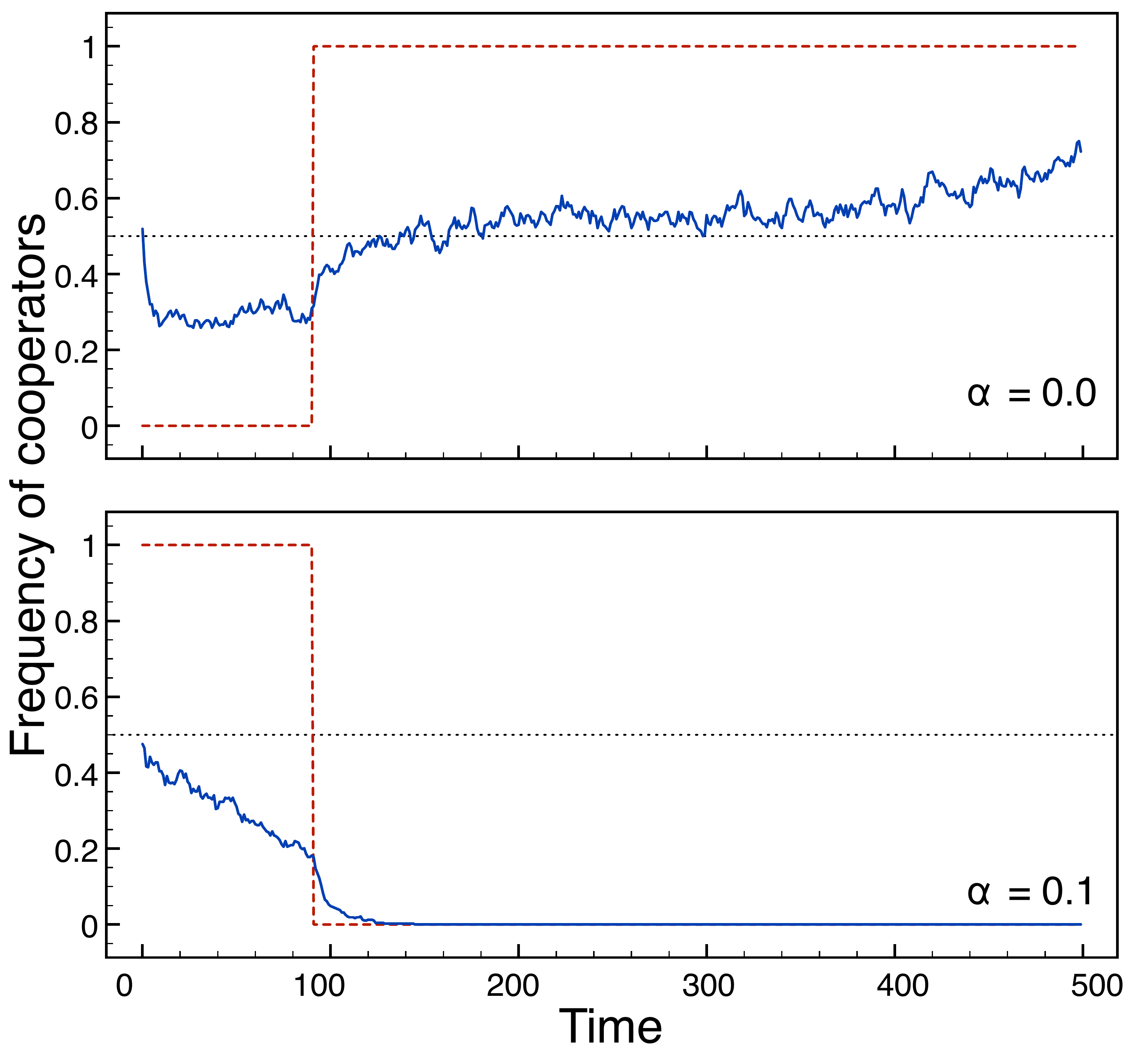}
}
\caption{(Color online) Evolution of cooperation around the most connected hub of a scale-free network with $\bar k = 8$. The game is a rescaled PD with $b=1.35$ for $\alpha = 0$ (upper panel) and $\alpha = 0.1$ (lower panel). The fraction of C neighbors is shown in solid lines and the strategy of the hub in dashed lines (D corresponds to 0; C to 1). As a reference, the level of 50\% cooperation is depicted in dotted lines. The most connected hub is initially set to D (upper panel) or to C (lower panel). The rest of the population is initialized to around 50\% Cs.}
\label{dyn_hub}
\end{figure}

Both Cs and Ds do better when they are surrounded by Cs.  By spreading defective behavior, hot Ds become less and less successful, since the number of their C neighbors decreases. Hot Cs, on the contrary, see their pay-off increased by spreading their own strategy. The more hot Cs are imitated the more they earn and the more difficult it is for a surrounding D to invade. A typical example of such ``hub dynamics'' is illustrated in Fig.~\ref{dyn_hub}~(upper panel) for the most connected hub of a scale-free network. The hub is D at the beginning of the simulation, while the rest of the population is initialized to around 50\% Cs. Many C neighbors imitate the defective hub (or other surrounding Ds) during the first steps of simulation, so that the proportion of C neighbors is reduced to approximately 30\%. As a consequence, the total pay-off of the hub is reduced, and the hub becomes vulnerable to invasion from a neighboring C. When the hub becomes a C, more and more of its D neighbors also switch their strategies. Consequently, the proportion of C neighbors (and the total pay-off to the hub) increases and is maintained at a high level afterwards. The presence of such positive feedback mechanism, and the fact that it only works for Cs, greatly enhances cooperation in degree-heterogeneous graphs and, particularly, in scale-free networks~\cite{Santos2006}. \\
The introduction of conformity decreases the bias in the flow of information in degree-heterogeneous graphs, making hubs vulnerable to invasion from their cold neighbors. While hubs are unlikely to imitate their low connected neighbors when using a pay-off biased rule, nothing prevents them from imitating a cold surrounding player if it holds the strategy of the local majority (see Fig.~\ref{fig:model}(d)). Since the fraction of Cs generally decreases at the outset of the simulation (see the first time steps of the curves shown in Fig.~\ref{dyn_sf}), conformity further favors Ds, which become predominant in the population. An example of this dynamics is shown in Fig.~\ref{dyn_hub}~(lower panel). Initially, the hub is a C. Many of the hub's neighbors turn to defection during the first time steps, making cooperation the less common strategy in the hub's neighborhood. Around the 100th time step, the hub imitates by conformity one of its defector neighbors, leading to a quicker decrease in the proportion of Cs in its neighborhood. Shortly after, Cs completely vanish around the most connected hub.
During those first time steps, hubs imitating according to a conformist bias will have many chances of becoming Ds. When Cs are not initially in the large majority, such initial asymmetry in the strategies of the hubs can account for the negative effects of conformity in the evolution of cooperation in scale-free networks. Conformity partly reverses the flow of information on degree-heterogeneous networks so that hubs no longer conduct the dynamics and instead quickly conform to the general trend of the whole population.

\section{Conclusions}
\label{concl}

To sum up, we have investigated the effects of conformity in the evolution of cooperation on regular one-dimensional lattices (rings) and scale-free networks. This was done by proposing an updating rule that is a stochastic average of the traditional local replicator dynamics, which models pay-off biased imitation, and a conformist biased rule of transmission favoring the most common variants around focal individuals. We explored rings and scale-free networks with different average degrees, as well as different values of the propensity to conform $\alpha$. Two games representing social dilemmas were studied: the rescaled versions of the PD, and the SG. In addition to Monte Carlo simulations, we also used an extended pair-dynamics model to predict the average fraction of cooperators in equilibrium, and compare them with the results obtained from our simulations.\\
The results presented in this paper show that whether conformity strengthens or weakens the evolution of cooperation depends on the intrinsic characteristics of the underlying graph. In the PD, conformity favors cooperation on rings by allowing clusters of Cs forming more easily. Conversely, it can hinder cooperation in scale-free networks for medium to large values of $b$, due to the exposure of hubs to the opinions of the local majority in their neighborhoods. In particular, and already for small amounts of conformity in the imitation rules of the players, scale-free networks do not show the great improvement over regular structures that has been previously reported in the literature. In the SG, conformity fosters cooperation on rings in the case $k=4$ for all values of the cost-to-benefit ratio $r$, and for low to medium values of $r$ in the case $k=8$. In scale-free networks, conformity is rather detrimental for large values of $r$. Thus, for both the PD and the SG, conformity often hinders the evolution of cooperation on scale-free networks for the cultural evolutionary dynamics described in this paper.\\
It is worth pointing out that other factors dismissing the advantage of scale-free networks in the evolution of cooperation have been identified, such as participation costs~\cite{Masuda2007}, other positive affine transformations of the pay-off matrix~\cite{Tomassini2006a,Luthi-Tom-Pest-09}, and the use of average instead of accumulated pay-offs~\cite{Santos2006b}. While these factors are extrinsic to the imitation rules of the agents, conformity is a simple mechanism undoubtedly present in our social learning psychology and central to better understand cultural dynamics and the way cooperation evolves on real social networks.\\

\section{Acknowledgments}
The authors would like to thank Yamir Moreno and Attila Szolnoki for useful advice and discussions. 
We are also grateful to the anonymous reviewers for their insightful comments. J. Pe\~na acknowledges financial support by the \emph{PERPLEXUS} project of the European Commission (grant IST-034632). E. Pestelacci and M. Tomassini acknowledge financial support by the Swiss National Science Foundation (grant 200020-119719/1). 

\section{Appendix: Pair approximation}

An analytical approximation of the dynamics of evolutionary games on graphs can be obtained by means of pair approximation~\cite{Matsuda1992,Baalen1998}. For detailed surveys of this technique, and its applications to games on graphs, we refer the interested reader to Refs.~\cite{Hauert2004, Hauert2005, Szabo2007}. We limit ourselves to briefly introduce the pair approximation and to explain how we have extended it for taking into account conformity in the imitation rules of the players.\\
Pair approximation is a method for constructing a system of ordinary differential equations for the global frequencies of strategies by tracking the changes in the frequencies of strategy pairs. In our case, we are interested in determining the global frequency $\rho$ of Cs by tracking the fluctuations in $p_{c,c}$, $p_{c,d}$, $p_{d,c}$ and $p_{d,d}$, where $p_{s,s'}$ is the probability of having an individual playing strategy $s$ connected to an individual playing strategy $s'$. For pair approximation to be consistent with the mean-field approach, it is assumed that $p_s = \sum_{s'} p_{s,s'}$. Furthermore, and in order to ``close'' the set of equations, configurations of triplets and more complicated configurations are approximated by the configuration probabilities of strategy pairs. For example, the configuration probability of the triplet $s,s',s''$ is approximated by $p_{s,s',s''}=p_{s,s'}p_{s',s''}/p_{s'}$. It is important to note that pair approximation (i) requires regular graphs and (ii) corrections arising from loops are ignored. 
Finally, note that the predictions of the pair approximation for any two regular graphs with the same degree $k$ are exactly the same. This allows us to compare our results to those of~\cite{Hauert2004} when $\alpha = 0$.
\begin{figure}[t]
\centerline{
	\includegraphics[width=2.0in,bb=0 0 393 323]{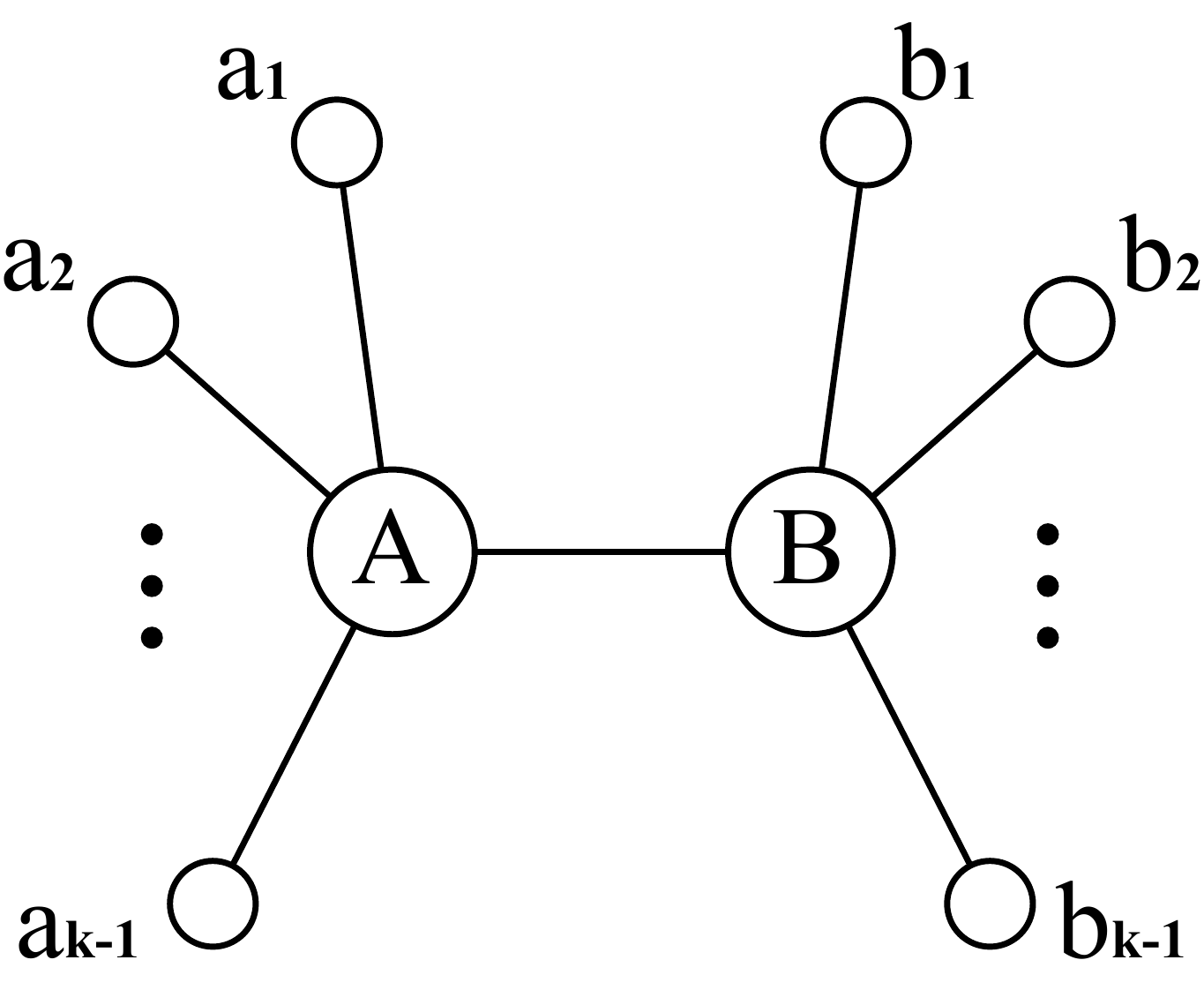}
}
\caption{A generic configuration for pair approximation. $A$ is the focal individual, $B$ is $A$'s cultural model, $a_1, a_2, \ldots, a_{k-1}$ are $A$'s neighbors other than $B$, and  $a_1, a_2, \ldots, a_{k-1}$ are $B$'s neighbors other than $A$. $A$ and $B$ are assumed to have no common neighbors, i.e. triangles and loops are neglected.}
\label{fig:pa}
\end{figure}   
\\
Let us consider individuals sitting on the vertices of a graph of degree $k$. Whenever a randomly chosen site $A$ updates its strategy, a random neighbor $B$ is selected as $A$'s cultural model. Common neighbors of any pair of vertices are considered to be independent by pair approximation (i.e. loops are neglected). Thus, let us denote by $a_1, \ldots, a_{k-1}$ (resp. $b_1, \ldots, b_{k-1}$) the $k-1$ the neighbors of $A$ (resp. $B$) other than $B$ (resp. $A$). The probability of a generic configuration (see Fig.~\ref{fig:pa}) is given by:
\begin{equation*}
p_{A,B}\frac{\prod_{i=1}^{k-1}p_{a_i,A}p_{b_i,B}}{p_A^{k-1}p_B^{k-1}}.
\end{equation*}
The probability that the pair $A,B$ becomes $B,B$ is calculated by multiplying the transition probability $\sigma_{A \rightarrow B}$ by the configuration probability and summing over all possible configurations, so that:
\begin{displaymath}
p_{A,B \rightarrow B.B} = \sum_{a_1,\ldots,a_{k-1}}\sum_{b_1,\ldots,b_{k-1}} \sigma_{A \rightarrow B} \times
p_{A,B}\frac{\prod_{i=1}^{k-1}p_{a_i,A}p_{b_i,B}}{p_A^{k-1}p_B^{k-1}} . 
\end{displaymath}
In our model, the transition probability $\sigma_{A \rightarrow B}$ depends not only on the payoffs of $A$ and $B$ but also on $\alpha$ (the probability to imitate according to a conformist bias) and on the number of players among $a_1, \ldots, a_{k-1}$ playing the same strategy of $A$ and $B$. The transition probability is given by:
\begin{eqnarray*}
\sigma_{A \rightarrow B} & = & (1-\alpha) f\left(\frac{\Pi_B(b_1,\ldots,b_{k-1})-\Pi_A(a_1,\ldots,a_{k-1})}{k\theta}\right) + \\
& & \alpha f \left(\frac{n_B(a_1,\ldots,a_{k-1},B)-n_A(a_1,\ldots,a_{k-1},B)}{k}\right) , 
\end{eqnarray*}
where $\Pi_B(x_1,\ldots,x_{k-1})$, $\Pi_A(x_1,\ldots,x_{k-1})$ denote the payoffs of $B$ ($A$) interacting with $x_1,\ldots,x_{k-1}$ plus $A$ ($B$), and $n_B(a_1, \ldots, a_{k-1},B)$, $n_A(a_1, \ldots, a_{k-1},B)$ specify the number of players with strategy $B$ ($A$) among $a_1,\ldots,a_{k-1}$ and $B$. The definitions of the parameter $\theta$ and the function $f$ are given in Section~\ref{sec:model}.\\
Whenever $A$ imitates $B$, the pair configuration probabilities change so that $p_{B,B}$, $p_{B,a_i},\ldots, p_{B,a_{k-1}}$ increase, while $p_{A,B}$, $p_{A,a_i},\ldots, p_{A,a_{k-1}}$ decrease. All these changes lead to a set of ordinary differential equations governing the dynamics of the system:
\begin{eqnarray*}
\dot{p}_{c,c} & = & \sum_{a_1,\ldots,a_{k-1}} \left(n_c(a_1, \ldots, a_{k-1})+1\right) \prod_{i=1}^{k-1}p_{d,a_i} \sum_{b_1,\ldots,b_{k-1}} \prod_{j=1}^{k-1}p_{c,b_j} \times \\
& &  \left\{ (1-\alpha) f \left(\frac{\Pi_c(b_1,\ldots,b_{k-1})-\Pi_d(a_1,\ldots,a_{k-1})}{k\theta}\right) + \alpha f \left(\frac{2 n_c(a_1, \ldots, a_{k-1})+2-k}{k}\right) \right\}  - \\
& & \sum_{a_1,\ldots,a_{k-1}} n_c(a_1, \ldots, a_{k-1})  \prod_{i=1}^{k-1}p_{c,a_i} \sum_{b_1,\ldots,b_{k-1}} \prod_{j=1}^{k-1}p_{d,b_j} \times \\
& & \left\{ (1-\alpha) f \left(\frac{\Pi_d(b_1,\ldots,b_{k-1})-\Pi_c(a_1,\ldots,a_{k-1})}{k\theta}\right) + \alpha f \left(\frac{k-n_c(a_1, \ldots, a_{k-1})}{k}\right) \right\}
\end{eqnarray*}
\begin{eqnarray*}
\dot{p}_{c,d} & = & \sum_{a_1,\ldots,a_{k-1}} \left(\frac{k}{2}-1-n_c(a_1,\ldots,a_{k-1})\right) \prod_{i=1}^{k-1}p_{d,a_i}  \sum_{b_1,\ldots,b_{k-1}} \prod_{j=1}^{k-1}p_{c,b_j} \times\\
& &  \left\{ (1-\alpha) f \left(\frac{\Pi_c(b_1,\ldots,b_{k-1})-\Pi_d(a_1,\ldots,a_{k-1})}{k\theta}\right) + \alpha f \left(\frac{2 n_c(a_1, \ldots, a_{k-1})+2-k}{k}\right) \right\}  - \\
& & \sum_{a_1,\ldots,a_{k-1}} \left(\frac{k}{2}-n_c(a_1, \ldots, a_{k-1})\right) \prod_{i=1}^{k-1}p_{c,a_i} \sum_{b_1,\ldots,b_{k-1}}  \prod_{j=1}^{k-1}p_{d,b_j} \times \\
& & \left\{ (1-\alpha) f \left(\frac{\Pi_d(b_1,\ldots,b_{k-1})-\Pi_c(a_1, \ldots, a_{k-1})}{k\theta}\right) + \alpha f \left(\frac{k-n_c(a_1, \ldots, a_{k-1})}{k}\right) \right\} , \\ 
\end{eqnarray*}
where $n_c(a_1, \ldots, a_{k-1})$ gives the number of Cs among $a_1, \ldots, a_{k-1}$ and $\Pi_c(x_1, \ldots, x_{k-1})$, $\Pi_d(x_1, \ldots, x_{k-1})$ denote the payoffs of a C (D) interacting with $x_1, \ldots, x_{k-1}$ plus a D (C). Because of the symmetry condition $p_{c,d} = p_{d,c}$ and the constraint $p_{c,c}+p_{c,d}+p_{d,c}+p_{d,d} = 1$ these two differential equations are sufficient to describe the system. Note that whenever $\alpha = 0$ the system of equations is equivalent to that derived in the supplementary information of Ref.~\cite{Hauert2004} and the appendix of Ref.~\cite{Hauert2005}. Following those works, the above equations also omit the common factor $2 p_{c,d}/(\rho^{k-1}p_d^{k-1})$, which has no influence in the equilibria of the system. The equilibrium values $\hat{p}_{c,c}$, $\hat{p}_{c,d}$, were obtained by numerically integrating the equations after specifying initial conditions for $10^{10}$ time steps. In all cases, $p_{c,c}(0)=(\rho(0))^2$, $p_{c,d}(0)=\rho(0)(1-\rho(0))$. The equilibrium frequency of Cs was then approximated by $\hat{p}_c = \hat{p}_{c,c}+\hat{p}_{c,d}$.

\end{document}